\documentclass{iopjournal}

\usepackage{amsmath}
\usepackage{amssymb}
\usepackage{amsfonts}
\usepackage{physics}
\usepackage{float}

\begin{document}

\articletype{Paper} 
\title{Emergent Quantum Droplets in Logarithmic Klein-Gordon Models of Bose-Einstein Condensates}

\author{Kevin Hernández$^1$$^*$\orcid{0000-0002-5739-3859} and Elías Castellanos$^2$\orcid{0000-0002-7615-5004} }

\affil{$^{1,}$Escuela de física, Universidad de El Salvador, 
 Final de Av. Mártires y Héroes del 30 de julio, San Salvador, El Salvador, América Central.}

\affil{$^2$Escuela de Ingenier\'ia y Ciencias, Tecnol\'ogico de Monterrey, Carr. Lago de Guadalupe Km. 3.5, Estado de M\'exico 52926, M\'exico.}




\begin{abstract}
We study a relativistic scalar field model for self-bound Bose-Einstein condensates (BECs) by analyzing a nonlinear Klein-Gordon equation with cubic and logarithmic interactions. This framework captures essential features of quantum droplets, such as self-trapping and finite energy configurations, which emerge from the interplay between attractive and repulsive terms. By performing the non-relativistic limit, we derive a generalized Gross-Pitaevskii equation with a logarithmic correction, consistent with recent models used to describe ultra-cold atomic gasses beyond mean-field theory. We construct the corresponding Lagrangian density, identify conserved quantities via Noether’s theorem, and compute the energy-momentum tensor. Numerical solutions of the BEC parameters are shown, establishing the foundations for a field-theoretic description of relativistic condensates with a logarithmic interaction. This model provides a unified approach to investigate relativistic effects in quantum droplets and enriches the theoretical landscape of Bose-Einstein condensates with non-standard interactions. The resulting dynamics exhibit stable oscillatory regimes consistent with self-bound condensate configurations.
\end{abstract}
\section{Introduction}
Since the first experimental realization of Bose–Einstein condensates (BEC) in dilute alkali gasses, these macroscopic quantum systems have become a cornerstone for studying collective quantum phenomena on a mesoscopic scale \cite{RevModPhys.71.463, wynar2000molecules, PhysRevLett.80.2027}. A BEC represents a phase of matter in which a large fraction of bosons occupies the same quantum ground state, giving rise to coherence and superfluidity described effectively by a macroscopic wave function. The dynamics and equilibrium properties of such condensates are well captured by the Gross–Pitaevskii equation (GPE), a nonlinear Schrödinger equation with a cubic self-interaction term arising from mean-field contact interactions. Despite its success, the standard GPE framework assumes weak interactions and neglects beyond-mean-field effects, which become crucial when the balance between attraction and repulsion is subtle or near collapse conditions. Nonlinear field equations play a crucial role in modern theoretical physics, appearing in contexts such as quantum field theory, condensed matter physics, and cosmology. Under specific conditions, quantum fluctuations and higher-order correlations can stabilize the condensate against collapse, leading to the emergence of Quantum (QD)— self-bound, liquid-like states emergent in ultracold atomic gases — reveal a striking interplay between quantum fluctuations and mean-field interactions 
\cite{luo2021new, cabrera2018quantum, PhysRevLett.116.215301, zloshchastiev2012volume, Zloshchastievql, Zloshchastievdilute}. While most theoretical descriptions of these droplets in Bose–Einstein condensates (BEC) have relied on extended Gross–Pitaevskii frameworks, including Lee–Huang–Yang (LHY) corrections \cite{dong2020multi, avdeenkov2011quantum, astrakharchik2018dynamics}, an alternative and conceptually illuminating approach is to study relativistic scalar-field models in which nonlinearity takes a special role \cite{visser2005massive, castellanos2013klein, megias2022nonlinear}.
Relativistic effects in Bose--Einstein condensation have been studied in detail in \cite{PhysRevLett.99.200406}, where the critical temperature of a relativistic ideal Bose gas was computed exactly, including the contribution of antibosons. The analysis shows that, at sufficiently high temperatures, pair production becomes relevant and the presence of antibosons lowers the Helmholtz free energy at the condensation point. This implies that neglecting antiboson contributions leads to a metastable description of the system. These results highlight the importance of relativistic corrections in bosonic systems and motivate the study of generalized relativistic models for Bose--Einstein condensates, particularly when nonlinear interactions are included. In this context, extensions of the Klein--Gordon framework provide a natural setting to explore effective interactions beyond the standard mean-field approximation.

In Ref. \cite{10.1143/PTP.115.1047, visser2005massive, castellanos2013klein}, Bose--Einstein condensation has also been explored in cosmological contexts, where it has been proposed as a possible mechanism underlying dark energy and structure formation. In such models, the condensation of a bosonic field can drive cosmic acceleration, while its dynamical instabilities may lead to the formation of localized structures, potentially associated with dark matter. These approaches, often based on relativistic extensions of the Gross--Pitaevskii equation, highlight the rich interplay between nonlinear dynamics, gravitational collapse, and collective quantum effects. This perspective further motivates the study of generalized relativistic condensate models, where additional nonlinear terms may play a significant role in the emergent dynamics.

The possibility that dark matter can be described as a Bose--Einstein condensate has been explored in Ref.~\cite{Böhmer_2007}. In this framework, galactic halo dynamics are modeled using the Gross--Pitaevskii equation, which, through the Madelung representation, yields a hydrodynamic description of a self-gravitating quantum fluid. For quartic interactions, the system obeys a polytropic equation of state with index $n=1$, and its structure can be described within the Thomas--Fermi approximation by the Lane--Emden equation. These models successfully reproduce galactic rotation curves and predict distinct gravitational lensing effects, supporting the relevance of Bose--Einstein condensates as dark matter candidates.

The free expansion of a Bose–Einstein condensate refers to the nonequilibrium dynamics that follow the sudden release of the atomic cloud from its external trapping potential. During this process, the condensate evolves under its internal interactions and kinetic pressure alone, revealing essential information about its coherence, interaction strength, and stability mechanisms. However, in systems exhibiting self-bound behavior—such as quantum droplets—the expansion can be dramatically suppressed or even entirely arrested due to the competing effects of attractive and repulsive nonlinearities. Numerical studies have shown that this self-stabilization manifests as oscillatory breathing modes, partial expansions, or metastable droplet remnants, depending on the interaction parameters and dimensionality of the system, in Ref. \cite{rodriguez2021oscillating}, we reported the emergence of quantum droplets composed of rubidium, sodium, and lithium atoms undergoing a cigar-shaped free expansion, in which the external trapping potential was released radially while preserved axially. This configuration led to the formation of stable, self-bound droplet states. We further extended this framework to model nonrelativistic bosonic stars as gravitational quantum droplets \cite{mastache2024non}, analyzing their radial free expansion under a partially confined potential. Additionally, we investigated the influence of rotation on the Bose–Einstein condensate \cite{hernandez2025rotating}, showing how angular momentum modifies the droplet’s morphology, stability, and expansion dynamics. The analysis of free expansion thus provides a powerful probe for understanding the underlying energy balance and the role of quantum correlations in self-bound quantum matter.

The logarithmic Klein–Gordon (LKG) model combines a relativistic kinetic term with a non-polynomial self-interaction that naturally encodes saturating, self-limiting behavior; these features make the LKG framework a promising minimal model for exploring the formation, stability, and dynamics of self-bound quantum droplets, although it has no analytical solution \cite{gorka2009logarithmic, bartkowski2008one, morris1978classical}.

In this work, we explore how a logarithmic nonlinearity in a Klein–Gordon-type equation gives rise to emergent droplet solutions that share the defining characteristics of experimentally observed quantum droplets: finite energy, self-confinement without external trapping, and robustness against collapse.

Because the LKG formalism admits a clear relativistic-to-nonrelativistic reduction, it provides a controlled pathway to connect droplet physics in condensates with field-theoretic insights about solitonic and self-bound states.

We present analytic and numerical evidence that LKG models support families of stable, finite-energy droplets for parameter ranges relevant to ultracold Bose gasses. After deriving the energy functional and conserved quantities, we analyze the existence and linear stability of localized solutions, characterize their scaling with particle number and interaction parameters, and study dynamical responses to perturbations.

Our results indicate that logarithmic nonlinearities capture key thermodynamic and dynamical aspects of quantum droplets while offering a compact, flexible framework for exploring extensions such as rotation, external potentials, and multi-component mixtures.
\section{Logarithmic Klein-Gordon Models of Bose-Einstein Condensates}
\subsection{Lorentz invariance and non-relativistic limit}
Relativistic effects on Bose--Einstein condensation have been studied in detail in early works such as Ref. \cite{Fujita1991}, where the condensation of free relativistic bosons was analyzed for both massive and massless cases. It was shown that the existence and nature of the phase transition strongly depend on the particle mass and the dimensionality of the system. In particular, for finite-mass bosons, condensation occurs only in three dimensions and the critical temperature is lower than in the corresponding nonrelativistic case. These results highlight that relativistic corrections can significantly modify the thermodynamic properties of bosonic systems, motivating the study of generalized relativistic models of Bose--Einstein condensates beyond the standard mean-field approximation. In this work, we analyze a relativistic scalar field equation that extends the $3+1$ Klein-Gordon equation by incorporating cubic and logarithmic nonlinearities as shown in Ref. \cite{gorka2009logarithmic}:

\begin{equation}
\left( \Box + \frac{m^2 c^2}{\hbar^2} \right) \Psi = \lambda |\Psi|^2 \Psi - \beta \ln\left(\alpha|\Psi|^2 \right) \Psi.
\label{EQ:LKG}
\end{equation}

The presence of the cubic term resembles self-interactions commonly found in quantum field theories, while the logarithmic term introduces a novel form of nonlinearity that has been studied in the context of effective models for Bose-Einstein condensates, gravity-inspired theories, and wave turbulence \cite{Zloshchastievql, zloshchastiev2012volume, Zloshchastievdilute}.

The given equation is relativistically invariant if $\Psi$ transforms as a relativistic scalar field. The d'Alembertian operator $\Box$ and the mass term $\frac{m^2 c^2}{\hbar^2} \Psi$ are naturally Lorentz-invariant. The nonlinear terms, $|\Psi|^2 \Psi$ and $\ln(\alpha |\Psi|^2) \Psi$, remain invariant if $\Psi$ is a Lorentz scalar since both $|\Psi|^2$ and its logarithm transform as scalars. Thus, the equation preserves its form under Lorentz transformations, confirming its relativistic invariance in the context of a scalar field theory.
To obtain the non-relativistic limit of the Eq. \ref{EQ:LKG}, we follow the standard fast-oscillating wave approximation of the Klein-Gordon equation.
In the non-relativistic limit, the wavefunction can be written as:

\begin{equation}
\Psi = e^{-i m c^2 t / \hbar} \psi(x,t),
\end{equation}

where $\psi(x,t)$ varies slowly compared to the rapid oscillations introduced by the mass term. In the non-relativistic limit, we neglect $\frac{1}{c^2} \frac{\partial^2 \psi}{\partial t^2}$, obtaining:

\begin{equation}
\Box \Psi \approx e^{-i m c^2 t / \hbar} \left( - \frac{m^2 c^2}{\hbar^2} \psi - \frac{2 i m}{\hbar} \frac{\partial \psi}{\partial t} - \nabla^2 \psi \right).
\end{equation}

Substituting into the equation:

\begin{equation}
- \frac{2 i m}{\hbar} \frac{\partial \psi}{\partial t} - \nabla^2 \psi = \lambda |\psi|^2 \psi - \beta \ln(\alpha |\psi|^2) \psi.
\end{equation}

Multiply by \(-\frac{\hbar^2}{2m}\):
\[
i \hbar \frac{\partial \psi}{\partial t} = -\frac{\hbar^2}{2m} \nabla^2 \psi -\frac{\hbar^2}{2m}\left( \lambda |\psi|^2 - \beta \ln(\alpha |\psi|^2) \right) \psi.
\]

Define:
\[
G \equiv -\frac{\hbar^2}{2m} \lambda, \quad B \equiv \frac{\hbar^2}{2m} \beta.
\]

Thus:
\[
i \hbar \frac{\partial \psi}{\partial t} = -\frac{\hbar^2}{2m} \nabla^2 \psi + \left( G |\psi|^2 + B \ln(\alpha |\psi|^2) \right) \psi.
\]

this equation is a nonlinear Schrödinger equation, similar to the Gross-Pitaevskii equation, but with cubic and logarithmic interaction terms. In the non-relativistic limit, the relativistic equation reduces to a Schrödinger equation with nonlinear interactions \cite{rodriguez2021oscillating, Zloshchastievql, Zloshchastievdilute}.

\subsection{Lagrangian of the BEC in LKG model}

In order to capture the self-trapping and stability mechanisms observed in logarithmic Bose–Einstein condensates, it is convenient to formulate the theory from a Lagrangian perspective as indicated in Ref.\cite{fagnocchi2010relativistic, 10.1143/PTP.115.1047}. The logarithmic nonlinearity naturally arises from a field-theoretic action where the potential term depends on the local density through a logarithmic function. Even though the model admits no exact analytical solution, the Lagrangian approach allows one to derive the corresponding field equations, conserved quantities, and energy functional in a systematic manner:

\begin{equation}
    \mathcal{L} = \partial_\mu \Psi^* \partial^\mu \Psi - \frac{m^2 c^2}{\hbar^2} |\Psi|^2 + \frac{\lambda}{2} |\Psi|^4 - \beta |\Psi|^2 \ln\left(\alpha |\Psi|^2 \right) + \beta |\Psi|^2
\label{Eq: DL-LKG}
\end{equation}
And applying the Euler-Lagrange equation:
\[
\frac{\partial \mathcal{L}}{\partial \Psi^*} - \partial_\mu \left( \frac{\partial \mathcal{L}}{\partial (\partial_\mu \Psi^*)} \right) = 0,
\]

we get Eq. \ref{EQ:LKG}, having defined the Lagrangian density of the logarithmic Klein–Gordon field, we proceed to identify the physical quantities that remain invariant under continuous transformations. Following Noether’s theorem, each symmetry of the Lagrangian corresponds to a conservation law: global phase invariance yields particle-number conservation, and spacetime translational invariance yields the conservation of energy and momentum. These conserved quantities form the foundation for analyzing equilibrium configurations and dynamical stability of the emergent quantum droplets.
naturally, The Lagrangian density defined in Eq. \ref{Eq: DL-LKG} is invariant under the global $U(1)$ phase transformation:
\begin{equation}
\Psi \rightarrow e^{i\theta}\Psi, \qquad 
\Psi^* \rightarrow e^{-i\theta}\Psi^*,
\end{equation}
which corresponds to the invariance of the theory under a global phase shift. According to Noether's theorem, this symmetry gives rise to a conserved four-current:
\begin{equation}
j^\mu = 
\frac{\partial \mathcal{L}}{\partial (\partial_\mu \Psi)} \, \delta \Psi
+
\frac{\partial \mathcal{L}}{\partial (\partial_\mu \Psi^*)} \, \delta \Psi^* .
\end{equation}

For the present Lagrangian, this yields
\begin{equation}
j^\mu = i \left( \Psi \, \partial^\mu \Psi^* - \Psi^* \, \partial^\mu \Psi \right),
\end{equation}
which satisfies the local conservation law
\begin{equation}
\partial_\mu j^\mu = 0.
\end{equation}

The conserved charge associated with the total particle number is obtained by integrating the time component of the current:
\begin{equation}
Q = \int d^3x \, j^0 = i \int d^3x \, \left( \Psi \, \partial^0 \Psi^* - \Psi^* \, \partial^0 \Psi \right).
\end{equation}
In order to find the canonical moment-tensor we use Eq. \ref{Eq: DL-LKG} and define the potential term as
\begin{equation}
V(|\Psi|^2) = \frac{m^2 c^2}{\hbar^2} |\Psi|^2 - \frac{\lambda}{2} |\Psi|^4 + \beta |\Psi|^2 \ln(\alpha |\Psi|^2) - \beta |\Psi|^2,
\label{EQ:potencialsimplificado}
\end{equation}
so that the Lagrangian can be rewritten in the compact form
\begin{equation}
\mathcal{L} = \partial_\mu \Psi^* \partial^\mu \Psi - V(|\Psi|^2).
\end{equation}

The canonical energy-momentum tensor, as derived from Noether’s theorem, is given by
\begin{equation}
T^{\mu\nu} = 
\frac{\partial \mathcal{L}}{\partial(\partial_\mu \Psi)} \, \partial^\nu \Psi 
+ 
\frac{\partial \mathcal{L}}{\partial(\partial_\mu \Psi^*)} \, \partial^\nu \Psi^* 
- \eta^{\mu\nu} \mathcal{L}.
\end{equation}

Using
\begin{equation}
\frac{\partial \mathcal{L}}{\partial(\partial_\mu \Psi)} = \partial^\mu \Psi^*, 
\qquad 
\frac{\partial \mathcal{L}}{\partial(\partial_\mu \Psi^*)} = \partial^\mu \Psi,
\end{equation}
we then obtain
\begin{equation}
T^{\mu\nu} = 
\partial^\mu \Psi^* \partial^\nu \Psi 
+ 
\partial^\mu \Psi \partial^\nu \Psi^* 
- \eta^{\mu\nu} \left( \partial_\rho \Psi^* \partial^\rho \Psi - V(|\Psi|^2) \right).
\end{equation}

This tensor encodes the local energy and momentum densities of the complex field and provides the foundation for analyzing conserved quantities and the dynamical evolution of self-bound configurations. We can find density energy with $T^{00}$:
\begin{equation}
T^{00} =|\dot{\Psi}|^2 + |\nabla \Psi|^2 + V(|\Psi|^2),
\end{equation}
and moment-density as $T^{0i}$:
\begin{equation}
T^{0i} = \dot{\Psi}^* \partial_i \Psi + \dot{\Psi} \partial_i \Psi^*     
\end{equation}
It is easy to demonstrate the conservation of $T^{\mu\nu}$, taking the divergence of the energy-momentum tensor, we obtain

\begin{align*}
\partial_\mu T^{\mu\nu} 
&= \partial_\mu \left( \partial^\mu \Psi^* \partial^\nu \Psi + \partial^\mu \Psi \partial^\nu \Psi^* \right) - \partial^\nu \mathcal{L} \\
&= (\partial_\mu \partial^\mu \Psi^*) \, \partial^\nu \Psi 
+ \partial^\mu \Psi^* \, \partial_\mu \partial^\nu \Psi 
+ (\partial_\mu \partial^\mu \Psi) \, \partial^\nu \Psi^* \\
&+ \partial^\mu \Psi \, \partial_\mu \partial^\nu \Psi^* 
- \partial^\nu \mathcal{L}.
\end{align*}

Since partial derivatives commute, the terms can be rearranged. If the field equations

\[
\partial_\mu \partial^\mu \Psi + \frac{\partial V}{\partial \Psi^*} = 0, \hspace{0.5cm}
\partial_\mu \partial^\mu \Psi^* + \frac{\partial V}{\partial \Psi} = 0,
\]

are satisfied, then each contribution cancels, yielding $ \partial_\mu T^{\mu\nu} = 0 $, as a direct consequence of the Euler–Lagrange equations. This result ensures the local conservation of energy and momentum for the complex scalar field governed by the nonlinear potential \(V(|\Psi|^2)\).

For a stationary state of the form
\[
\Psi(\mathbf{r}, t) = \psi(\mathbf{r}) e^{-i \mu t / \hbar},
\]
we obtain
\[
\dot{\Psi} = -i \frac{\mu}{\hbar} \Psi \quad \Rightarrow \quad 
|\dot{\Psi}|^2 = \left( \frac{\mu}{\hbar} \right)^2 |\psi|^2.
\]
Hence, the corresponding energy density is given by
\[
T^{00} = \left( \frac{\mu}{\hbar c} \right)^2 |\psi|^2 + |\nabla \psi|^2 + V(|\psi|^2),
\]
and the total energy of the configuration reads
\[
E = \int d^3x \, 
\left[ 
\left( \frac{\mu}{\hbar c} \right)^2 |\psi|^2 
+ 
|\nabla \psi|^2 
+ 
V(|\psi|^2) 
\right].
\]

This expression represents the total energy of a stationary self-bound configuration characterized by the chemical potential $\mu$ and the nonlinear potential term $V(|\psi|^2)$.
\subsection{Stability analysis via chemical potential}
For a spherically symmetric quantum droplet, at the center we have:
\begin{itemize}
    \item The gradient vanishes: \(\nabla \psi = 0\),
    \item There are no cross-derivative terms.
\end{itemize}

Therefore, the stress tensor reduces to

\[
T^{ij}_{\text{center}} 
= 
\delta^{ij} 
\left[
\left( \frac{\mu}{\hbar c} \right)^2 |\psi(0)|^2 
- V(|\psi(0)|^2)
\right].
\]

We can interpret this quantity as an effective pressure:

\[
P = 
\left( \frac{\mu}{\hbar c} \right)^2 |\psi(0)|^2 
- V(|\psi(0)|^2)
\]

where the potential term is given by Eq. \ref{EQ:potencialsimplificado}.
If \(P > 0\), the droplet possesses an internal pressure and tends to expand;  
if \(P < 0\), an inward tension arises and the droplet contracts.  
The equilibrium condition \(P = 0\) thus determines the critical size or density of a self-sustained quantum droplet. Assuming a Gaussian ansatz:
\[
\psi(\mathbf{r}) = A e^{-r^2/(2a^2)}, \quad \text{with} \quad A^2 = \frac{N}{\pi^{3/2} a^3}
\]

At the center of the droplet, \( r = 0 \), we have \( |\psi(0)|^2 = A^2 \).  
Thus, the equilibrium condition \( P = 0 \) leads to

\[
\left( \frac{\mu}{\hbar c} \right)^2 A^2 
= V(A^2).
\]

Dividing both sides by \( A^2 \), we obtain

\[
\left( \frac{\mu}{\hbar c} \right)^2 
= \frac{V(A^2)}{A^2}
= 
\frac{m^2 c^2}{\hbar^2}
- \frac{\lambda}{2} A^2
+ \beta \ln(\alpha A^2)
- \beta,
\]

substituting 
\(
A^2 = \dfrac{N}{\pi^{3/2} a^3},
\)
we find

\[
\left( \frac{\mu}{\hbar c} \right)^2 
= 
\frac{m^2 c^2}{\hbar^2}
- \frac{\lambda N}{2\pi^{3/2} a^3}
+ \beta \ln\!\left( \alpha \frac{N}{\pi^{3/2} a^3} \right)
- \beta.
\]

and then:

\[
\mu = m c^2
\sqrt{
1
- \frac{\hbar^2}{m^2 c^2}
\left[
\frac{\lambda N}{2\pi^{3/2} a^3}
- \beta \ln\!\left( \alpha \frac{N}{\pi^{3/2} a^3} \right)
+ \beta
\right]}.
\]

For $\lambda =0$ and $\beta=0$ we obtain $\mu = mc^2$ and in the non-Relativistic limit we can apply $\qty(1+x)^m \approx 1+mx$ and get $
\mu \approx m c^2 + \mu_{\rm NR}$ with
\begin{equation}
\mu_{\rm NR} = - \frac{\hbar^2}{2 m} 
\left[
\frac{\lambda N}{2\pi^{3/2} a^3}
- \beta \ln\!\left( \alpha \frac{N}{\pi^{3/2} a^3} \right)
+ \beta
\right]. 
\label{Eq:muRela}
\end{equation}
The chemical potential $\mu$ plays a central role in identifying self-bound states in Bose--Einstein condensates. In Ref.~\cite{PhysRevLett.99.200406}, the chemical potential is determined from the thermodynamic properties of a relativistic Bose gas, taking into account the contribution of antibosons at finite temperature. In the limit $\beta = 0$, corresponding to the absence of logarithmic corrections in the present model, our results are consistent with the standard relativistic description reported in that work.. Unlike the total energy, which scales with the number of particles, the chemical potential measures the change in energy associated with adding or removing a particle, making it particularly suitable for systems with fixed particle number. By analyzing $\mu$ as a function of the condensate size or density, one can locate minima corresponding to stable configurations where attractive and repulsive interactions balance each other. These minima indicate self-confinement, as the system naturally favors the associated density profile. Thus, $\mu$ provides a direct and practical tool to detect and characterize autoconfined states. The minimum of the chemical potential, $\mu_{\rm NR}(a)$, indicates the equilibrium size of a self-bound state in a Bose-Einstein condensate. It corresponds to the point where the attractive and repulsive contributions balance, resulting in a stable configuration. Mathematically, the minimum is obtained by differentiating the chemical potential with respect to the condensate width, $a$, and setting the derivative to zero:
\[
\frac{d \mu_{\rm NR}}{da} = 0.
\]
Setting the derivative to zero for the minimum:

\[
- \frac{3 \lambda N}{2 \pi^{3/2} a^4} + \frac{3 \beta}{a} = 0 \quad \Rightarrow \quad a^3 = \frac{\lambda N}{2 \pi^{3/2} \beta}.
\]

Thus, the equilibrium width $a$ of the self-confined state is determined explicitly by the system parameters $\lambda$, $\beta$, and $N$. Thus, the chemical potential at the self-confined equilibrium state is
\[
\mu_{\rm NR}(a_0) = - \frac{\hbar^2 \beta}{2 m} \left[ 2 - \ln\left( \frac{2 \alpha \beta}{\lambda} \right) \right].
\]
In the analysis of self-bound states, it is important to distinguish between fixed and free parameters. Fixed parameters, such as the atomic mass $m$, the Planck constant $\hbar$, and the interaction strength $\lambda$ determined by the scattering length $a_s$, are intrinsic to the physical system and cannot be changed experimentally. In contrast, free parameters, including the logarithmic interaction strength $\beta$ and the scale factor $\alpha$, can be varied to explore different regimes of stability and confinement. By tuning these free parameters, one can study how the equilibrium width and chemical potential of the condensate respond, while the fixed constants ensure consistency with the underlying physical properties.

\begin{figure}[htbp!]
    \centering
\caption{Contour plot of the non-relativistic chemical potential $\mu_{\rm NR}(a, N)$ from Eq. \ref{Eq:muRela} for $^{87}$Rb, with $\beta = 1$ and $\alpha = 1 \times 10^{-15}$. The red line indicates the $\mu_{\rm NR} = 0$ contour, separating self-bound states ($\mu_{\rm NR} < 0$) from unbound configurations ($\mu_{\rm NR} > 0$). The plot illustrates how the equilibrium width of the condensate depends on the number of particles $N$ and highlights the parameter regime where self-confinement occurs.}
    \label{fig:placeholder1}
\end{figure}

Although the chemical potential $\mu$ provides useful information about the energetic balance of a Bose–Einstein condensate or a quantum droplet, it is not sufficient to determine the existence of self-bound states in general regions of the parameter space. The reason is that $\mu$ only reflects the energy per particle in a stationary configuration, but it does not account for the full stability conditions arising from spatial variations of the density or from collective excitations. In particular, the formation of quantum droplets requires a delicate balance between attractive and repulsive interactions, which is encoded in the total energy functional and its minimization, not solely in the value of $\mu$. Therefore, a negative or vanishing chemical potential may indicate binding tendencies, but it does not guaranty the existence or stability of a droplet without a full dynamical or variational analysis.
\section{Methodology: Relativistic Bose-Einstein Condensate}

We start from the nonlinear Klein-Gordon equation for a Bose-Einstein condensate from Ref. \cite{fagnocchi2010relativistic} and Eq. \ref{EQ:LKG} with $\beta=0$:

\begin{equation}
\left(\Box + \frac{m^2 c^2}{\hbar^2} \right) \Psi(\mathbf{r},t) = \lambda |\Psi(\mathbf{r},t)|^2 \Psi(\mathbf{r},t),
\end{equation}

and aim to study the temporal evolution of the condensate width $a(t)$ using a Gaussian variational ansatz. Assume a simple spherically symmetric Gaussian:

\begin{equation}
\Psi(\mathbf{r},t) =\frac{N^{1/2}}{\pi^{3/4} a^{3/2}} \exp\Big[-\frac{r^2}{2 a^2}\Big] e^{i \theta(t)},
\end{equation}

where:

\begin{itemize}
    \item $a(t)$ is the condensate width,
    \item $\theta(t)$ is the global phase.
\end{itemize}
The wave function is normalized to the total number of particles $N$:$
\int |\Psi(\mathbf{r},t)|^2 \, d^3r = N$.

\begin{figure}[htbp!]
    \centering
    \includegraphics[width=0.8\linewidth]{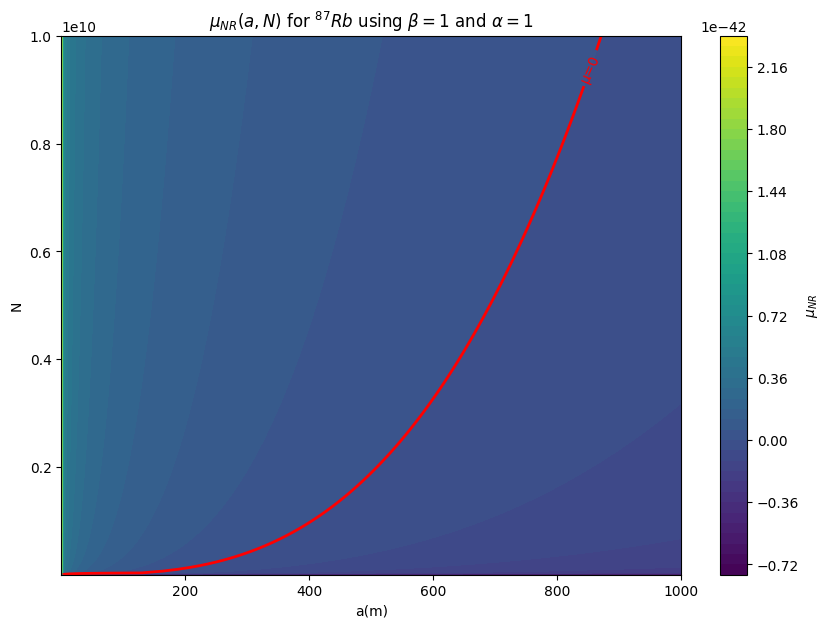}
    \caption{Same that Fig. \ref{fig:placeholder1} with $\beta = 1$ and $\alpha = 1$. As the number of atoms increases, a larger radius is required for the quantum droplet.}
    \label{fig:placeholder2}
\end{figure}
\subsection*{Effective Lagrangian}

The Lagrangian density is:

\begin{equation}
\mathcal{L} = \frac{1}{c^2} |\partial_t \Psi|^2 - |\nabla \Psi|^2 - \frac{m^2 c^2}{\hbar^2} |\Psi|^2 - \frac{\lambda}{2} |\Psi|^4
\end{equation}

Integrating over space with the Gaussian ansatz, we obtain the effective Lagrangian $L_{\text{eff}} = \int \mathcal{L} \, d^3r$, after integrating the Lagrangian density over all space, the effective Lagrangian becomes:
\begin{equation}
 L_{\rm eff} \;=\; \frac{N}{c^2}\,\dot\theta^{\,2} \;+\; \frac{3N}{2c^2}\,\frac{\dot a^{\,2}}{a^2} \;-\; \frac{3N}{2}\,\frac{1}{a^2} \;-\; N\,\frac{m^2 c^2}{\hbar^2} \;-\; \frac{\lambda\,N^2}{2^{5/2}\,\pi^{3/2}\,a^3}\,. 
\end{equation}

Since \(L_{\rm eff}\) does not depend on \(\theta\) explicitly, the conjugate momentum is conserved:
\[
\frac{d}{dt}\!\left(\frac{\partial L_{\rm eff}}{\partial\dot\theta}\right)=0
\quad\Longrightarrow\quad
\frac{\partial L_{\rm eff}}{\partial\dot\theta}=\frac{ N}{c^2}\dot\theta = \mbox{const} \equiv C.
\]
For a stationary state, we identify $\mu = \hbar \Omega$ as the chemical potential, giving $\dot{\theta} = \frac{\mu}{\hbar}$. This relation reflects the Goldstone mode associated with the broken $U(1)$ symmetry in the condensed phase. The constant phase rotation corresponds to a conserved particle number, with $\mu$ representing the energy cost to add one particle to the system.

Using the Euler--Lagrange equation for the variable $a(t)$

\[
\frac{d}{dt}\left(\frac{\partial L}{\partial \dot a}\right)
-
\frac{\partial L}{\partial a}
=0,
\]

we obtain the equation of motion

\begin{equation}
\ddot a
=
\frac{\dot a^2}{a}
+
\frac{c^2}{a}
+
\frac{\lambda N c^2}{2^{5/2}\pi^{3/2}a^2}.
\end{equation}

An interesting feature of this equation is the existence of an equilibrium radius $a_0$, obtained from the conditions $\dot a(0)=0$ and $\ddot a(0)=0$. This leads to

\begin{equation}
a_0
=
-\frac{\lambda N}{2^{5/2}\pi^{3/2}}.
\end{equation}

For the radius to have a physical meaning ($a_0>0$), the interaction parameter must satisfy $\lambda < 0$, which corresponds to an repulsive interaction. This result indicates that additional physical mechanisms are required to stabilize the condensate beyond the mean-field interaction.

The generalized momentum associated with $a$ is

\[
p_a
=
\frac{\partial L}{\partial \dot a}
=
\frac{3N}{c^2}\frac{\dot a}{a^2}.
\]

The conserved Hamiltonian of the system is

\begin{equation}
H
=
\frac{3N}{2c^2}\frac{\dot a^2}{a^2}
+
\frac{3N}{2a^2}
+
\frac{m^2c^2N}{\hbar^2}
+
\frac{\lambda N^2}{\pi^{3/2}2^{5/2}a^3}
=
E,
\end{equation}

where $E$ is a constant determined by the initial conditions

\[
a(0)=a_0, \qquad \dot a(0)=v_0.
\]

Thus,

\begin{equation}
E
=
\frac{3N}{2c^2}\frac{v_0^2}{a_0^2}
+
\frac{3N}{2a_0^2}
+
\frac{m^2c^2N}{\hbar^2}
+
\frac{\lambda N^2}{\pi^{3/2}2^{5/2}a_0^3}.
\end{equation}
and:
\[
\alpha = \frac{3N}{2c^2}, \qquad
\beta = \frac{3N}{2}, \qquad
\gamma = \frac{\lambda N^2}{\pi^{3/2}2^{5/2}}, \qquad
\delta = \frac{m^2c^2N}{\hbar^2}
\]

From the energy conservation:
\[
\alpha \frac{\dot{a}^2}{a^2} = E - \delta - \frac{\beta}{a^2} - \frac{\gamma}{a^3}
\]
\[
\dot{a}^2 = \frac{E - \delta}{\alpha} a^2 - \frac{\beta}{\alpha} - \frac{\gamma}{\alpha a}
\]

Separating variables:
\[
\frac{da}{dt} = \pm \sqrt{ \frac{E - \delta}{\alpha} a^2 - \frac{\beta}{\alpha} - \frac{\gamma}{\alpha a} }
\]
\[
dt = \pm \frac{da}{\sqrt{ \frac{E - \delta}{\alpha} a^2 - \frac{\beta}{\alpha} - \frac{\gamma}{\alpha a} }}
\]

Multiplying numerator and denominator by $\sqrt{\alpha}$:
\[
dt = \pm \frac{\sqrt{\alpha} \, da}{\sqrt{ (E - \delta) a^2 - \beta - \frac{\gamma}{a} }}
\]

Rewriting the denominator:
\[
(E - \delta) a^2 - \beta - \frac{\gamma}{a} = \frac{(E - \delta) a^3 - \beta a - \gamma}{a}
\]

Thus:
\[
dt = \pm \frac{\sqrt{\alpha} \sqrt{a} \, da}{\sqrt{ (E - \delta) a^3 - \beta a - \gamma }}
\]

Integrating both sides:

\begin{equation}
t - t_0 = \pm \sqrt{\alpha} \int \frac{\sqrt{a} \, da}{\sqrt{ (E - \delta) a^3 - \beta a - \gamma }}
\label{Eq:t-eliptic}
\end{equation}

presents several important physical and mathematical features:

\begin{itemize}
    \item \textbf{Elliptic Nature}: The integral is generally elliptic since the denominator contains a cubic polynomial in $a$ under a square root. This means the solution $a(t)$ can be expressed in terms of Weierstrass elliptic function $\wp(t)$ or Jacobi elliptic functions.

    \item \textbf{Turning Points}: The roots of the cubic polynomial $(E - \delta)a^3 - \beta a - \gamma = 0$ determine the \textbf{turning points} where $\dot{a} = 0$. These correspond to the minimum and maximum condensate widths during oscillation, or to collapse points for attractive interactions.

    \item \textbf{Physical Regimes}: Depending on the parameters:
    \begin{itemize}
        \item For $E - \delta > 0$ and dominant positive terms: \textbf{monotonic expansion}
        \item For bounded motion between two real roots: \textbf{oscillatory behavior} 
        \item For attractive interactions ($\gamma > 0$): possible \textbf{finite-time collapse}
    \end{itemize}

    \item \textbf{Special Cases}: The integral simplifies in certain limits:
    \begin{itemize}
        \item No interactions ($\gamma = 0$): reduces to elementary functions
        \item Non-relativistic limit ($c \to \infty$): reproduces Gross-Pitaevskii dynamics
        \item Static solution: when the cubic has a double root
    \end{itemize}

    \item \textbf{Numerical Evaluation}: For general parameters, the integral must be evaluated numerically, providing $t(a)$ inversely, which can then be inverted to obtain $a(t)$.
\end{itemize}

Eq. \ref{Eq:t-eliptic} provides a complete qualitative understanding of the condensate width dynamics without requiring explicit solution of the differential equation. In section \ref{section:bec-qd} we will show important results about the simulation.

\section{Time Evolution of the BEC Width with Cubic and Logarithmic Nonlinearities}
\label{section:bec-qd}

We consider a relativistic nonlinear equation describing a Bose--Einstein condensate with both cubic and logarithmic interactions as Eq. \ref{EQ:LKG}, where $\lambda$ characterizes the usual cubic mean--field interaction, while $\beta$ and $\alpha$ determine the strength of the logarithmic nonlinearity. Such logarithmic corrections can effectively model beyond--mean--field contributions and self--stabilizing mechanisms in quantum fluids.

\subsection{Gaussian Variational Ansatz}

To study the collective dynamics of the condensate, we employ a spherically symmetric Gaussian ansatz as REf. \cite{rodriguez2021oscillating, castellanos2013klein, hernandez2025rotating}, therefore,

\begin{equation}
\Psi(\mathbf r,t)
=
\frac{\sqrt{N}}{(\pi a^2)^{3/4}}
\exp\!\left(-\frac{r^2}{2a^2(t)}\right)
e^{i\theta(t)},
\end{equation}

where

\begin{itemize}
\item $a(t)$ represents the time--dependent condensate width,
\item $\theta(t)$ is the global phase,
\item $N$ is the total number of particles.
\end{itemize}

The wave function satisfies the normalization condition

\begin{equation}
\int |\Psi(\mathbf r,t)|^2 d^3 r = N.
\end{equation}

\subsection{Lagrangian Density}

The field equation can be derived from the Lagrangian density

\begin{equation}
\mathcal L
=
\frac{1}{2}\partial_\mu \Psi^* \partial^\mu \Psi
-
\frac{m^2 c^2}{2\hbar^2}|\Psi|^2
-
\frac{\lambda}{2}|\Psi|^4
+
\beta |\Psi|^2 \ln(\alpha |\Psi|^2).
\end{equation}

The effective Lagrangian governing the collective variables is obtained by integrating over space.

\begin{equation}
L_{\mathrm{eff}}(a,\dot a,\theta,\dot\theta)
=
\int \mathcal L \, d^3r ,
\end{equation}





the effective Lagrangian takes the form

\begin{equation}
L_{\mathrm{eff}}
=
\frac{3N}{4c^2}\dot a^2
+
\frac{Na^2}{2c^2}\dot\theta^2
-
V_{\mathrm{eff}}(a),
\end{equation}

where the effective potential reads

\begin{equation}
V_{\mathrm{eff}}(a)
=
\frac{3N\hbar^2}{4m^2 a^2}
+
\frac{3Nm^2 c^2}{4\hbar^2} a^2
+
\frac{\lambda N^2}{2(2\pi)^{3/2} a^3}
-
\beta N
\ln\!\left(
\frac{\alpha N}{(\pi a^2)^{3/2}}
\right).
\end{equation}

The different contributions have clear physical interpretations: the first term corresponds to the quantum pressure, the second arises from the relativistic mass term, the third describes the cubic interaction energy, and the last term represents the logarithmic correction.

\subsection{Phase Dynamics}

The Euler--Lagrange equation for the phase $\theta$ yields

\begin{equation}
\frac{d}{dt}
\left(
\frac{\partial L_{\mathrm{eff}}}{\partial \dot\theta}
\right)
=0,
\end{equation}

which leads to the conserved quantity

\begin{equation}
a^2 \dot\theta = Q,
\end{equation}

where $Q$ is a constant related to the global $U(1)$ symmetry of the system.

\subsection{Equation of Motion for the Condensate Width}

Applying the Euler--Lagrange equation to the variable $a(t)$,

\begin{equation}
\frac{d}{dt}
\left(
\frac{\partial L_{\mathrm{eff}}}{\partial \dot a}
\right)
-
\frac{\partial L_{\mathrm{eff}}}{\partial a}
=
0,
\end{equation}

we obtain the following ordinary differential equation governing the condensate width,

\begin{equation}
\ddot a
=
\frac{\hbar^2}{m^2 a^3}
+
\frac{Q^2}{a^3}
-
\frac{m^2 c^4}{\hbar^2} a
-
\frac{3\lambda N}{2(2\pi)^{3/2} a^4}
+
\frac{3\beta c^2}{a}.
\label{Eq:003}
\end{equation}

This equation describes the collective radial dynamics of the condensate. It contains several contributions with clear physical interpretations. The first
term, proportional to $1/a^3$, represents the quantum pressure originating
from the Heisenberg uncertainty principle. As the condensate becomes more
localized (smaller $a$), the kinetic energy increases, generating a repulsive
effect that tends to expand the system.

The second contribution, also proportional to $1/a^3$, arises from the
conserved quantity $Q$ associated with the global phase symmetry of the
system. This term acts as an additional effective kinetic pressure and
contributes to the stabilization of the condensate width.

The linear term proportional to $a$ originates from the relativistic mass
contribution in the Klein--Gordon description. This term acts as an effective
harmonic restoring force that tends to confine the condensate around an
equilibrium configuration.

The term proportional to $1/a^4$ corresponds to the cubic mean-field
interaction. Its sign depends on the interaction parameter $\lambda$. For
attractive interactions ($\lambda < 0$), this term favors compression of the
condensate and may lead to collapse if not counterbalanced by other
mechanisms.

Finally, the term proportional to $1/a$ arises from the logarithmic
nonlinearity. This contribution acts as an effective pressure that modifies
the equilibrium structure of the condensate and can provide an additional
stabilizing mechanism in regimes where attractive interactions dominate.
\subsection{Consistency with the Heisenberg Uncertainty Principle}

In previous works \cite{rodriguez2021oscillating, hernandez2025rotating, mastache2024non}, it has been emphasized that the effective equation governing the condensate width must be consistent with fundamental quantum mechanical constraints. In particular, the term responsible for the quantum pressure can be understood directly from the Heisenberg uncertainty principle,

\begin{equation}
\Delta x\,\Delta p \gtrsim \frac{\hbar}{2}.
\end{equation}

If the condensate has a characteristic spatial extension given by the variational parameter $a(t)$, we may estimate

\[
\Delta x \sim a .
\]

From the uncertainty relation, the corresponding momentum uncertainty is therefore

\begin{equation}
\Delta p \sim \frac{\hbar}{a}.
\end{equation}

The minimal kinetic energy associated with this momentum spread is

\begin{equation}
E_k \sim \frac{(\Delta p)^2}{2m}
\sim
\frac{\hbar^2}{2m a^2}.
\end{equation}

For a condensate containing $N$ particles, the total kinetic contribution scales as

\begin{equation}
E_k \sim \frac{N\hbar^2}{2m a^2}.
\end{equation}

This expression has the same functional dependence as the quantum pressure term obtained from the variational calculation. In particular, the corresponding effective force acting on the width can be estimated as

\begin{equation}
F(a) = -\frac{dE_k}{da}
\sim
\frac{N\hbar^2}{m a^3}.
\end{equation}

Dividing by an effective mass of order $Nm$, the resulting acceleration scales as

\begin{equation}
\ddot a \sim \frac{\hbar^2}{m^2 a^3}.
\end{equation}

This behavior coincides with the first term appearing in the dynamical equation for the condensate width as Eq. \ref{Eq:003}. Therefore, the quantum pressure term proportional to $1/a^3$ is a direct manifestation of the Heisenberg uncertainty principle. This provides an important physical consistency check for the variational description, ensuring that the condensate cannot collapse indefinitely due to the fundamental quantum mechanical limit on spatial localization.
\subsection{Approximate Equilibrium Radius of the Self-Bound Condensate}

An important property of the system is the existence of a stationary configuration in which the condensate width remains constant in time. Such equilibrium states are obtained from the dynamical equation for the width by imposing the conditions

\[
\dot a = 0, \qquad \ddot a = 0 .
\]

Applying these conditions to the Eq. \ref{Eq:003}, we obtain the equilibrium condition

\begin{equation}
\left(\frac{\hbar^2}{m^2}+Q^2\right)a_0
-
\frac{m^2 c^4}{\hbar^2}a_0^5
-
\frac{3\lambda N}{2(2\pi)^{3/2}}
+
3\beta c^2 a_0^3
=0 .
\end{equation}

In general this equation is highly nonlinear and does not admit a simple analytic solution. However, useful insight can be obtained by considering the regime in which the dominant balance occurs between the attractive cubic interaction and the repulsive logarithmic contribution. Neglecting the smaller terms, the equilibrium condition reduces approximately to

\begin{equation}
\frac{3\beta c^2}{a_0}
\approx
\frac{3|\lambda| N}{2(2\pi)^{3/2} a_0^4}.
\end{equation}

Solving for the equilibrium width yields the approximate relation

\begin{equation}
a_0^3
=
\frac{|\lambda|N}{2(2\pi)^{3/2}\beta c^2},
\end{equation}

or equivalently,

\begin{equation}
a_0
=
\left(
\frac{|\lambda|N}{2(2\pi)^{3/2}\beta c^2}
\right)^{1/3}.
\end{equation}

This result shows that the logarithmic nonlinearity provides an effective repulsive pressure that counterbalances the attractive cubic interaction, allowing the formation of a self-bound condensate. The equilibrium radius scales with the particle number as

\[
a_0 \propto N^{1/3},
\]

which is characteristic of self-stabilized quantum droplets \cite{rodriguez2021oscillating, pethick2008bose, mastache2024non, hernandez2025rotating, RevModPhys.71.463}. Consequently, the logarithmic interaction plays a crucial role in preventing collapse and stabilizing the condensate even in the presence of attractive mean-field interactions \cite{rodriguez2021oscillating, hernandez2025rotating, Zloshchastievql}.
\begin{table}[h]
\centering
\begin{tabular}{c c c}
\hline
Parameter & Symbol & Value \\
\hline
Atomic mass & $m$ & $1.443\times10^{-25}$ kg \\
Reduced Planck constant & $\hbar$ & $1.054\times10^{-34}$ J s \\
Speed of light & $c$ & $2.998\times10^{8}$ m/s \\
Scattering length & $a_s$ & $5.29\times10^{-9}$ m \\
Interaction strength & $g=4\pi\hbar^2 a_s/m$ & $5.1\times10^{-51}$ J m$^3$ \\
\hline
\end{tabular}
\caption{Physical constants for $^{87}$Rb used in the simulations.}
\label{Table1}
\end{table}

\begin{table}[h]
\centering
\begin{tabular}{c c c c c c}
\hline
Case & $N$ & $\lambda$ (J m$^3$) & $\beta$ (J) & $a(0)$ (m) & $\dot a(0)$ (m/s) \\
\hline
A & $5\times10^4$ & $5.1\times10^{-51}$ & $1\times10^{-32}$ & $3\times10^{-6}$ & $0$ \\

B & $1\times10^5$ & $5.1\times10^{-51}$ & $3\times10^{-32}$ & $2.5\times10^{-6}$ & $0$ \\

C & $2\times10^5$ & $-5.1\times10^{-51}$ & $5\times10^{-32}$ & $2\times10^{-6}$ & $0$ \\

D & $5\times10^5$ & $-5.1\times10^{-51}$ & $8\times10^{-32}$ & $1.5\times10^{-6}$ & $1\times10^{-4}$ \\

\hline
\end{tabular}
\caption{Representative parameters for $^{87}$Rb condensates used as input for the fourth--order Runge--Kutta (RK4) simulations of the width evolution equation.}
\label{Table2}
\end{table}

\begin{table}[h]
\centering
\begin{tabular}{c c c}
\hline
Parameter & Symbol & Value \\
\hline
Scattering length & $a_s$ & $2.75\times10^{-9}$ m \\
Interaction strength & $g=4\pi\hbar^2 a_s/m$ & $1.0\times10^{-50}$ J m$^3$ \\
\hline
\end{tabular}
\caption{Physical constants for $^{23}$Na used in the simulations.}
\label{table3}
\end{table}

\begin{table}[h]
\centering
\begin{tabular}{c c c c c c}
\hline
Case & $N$ & $\lambda$ (J m$^3$) & $\beta$ (J) & $a(0)$ (m) & $\dot a(0)$ (m/s) \\
\hline
A & $4\times10^4$ & $1.0\times10^{-50}$ & $8\times10^{-33}$ & $4\times10^{-6}$ & $0$ \\

B & $8\times10^4$ & $1.0\times10^{-50}$ & $2\times10^{-32}$ & $3\times10^{-6}$ & $0$ \\

C & $1.5\times10^5$ & $-1.0\times10^{-50}$ & $4\times10^{-32}$ & $2.5\times10^{-6}$ & $0$ \\

D & $3\times10^5$ & $-1.0\times10^{-50}$ & $6\times10^{-32}$ & $2\times10^{-6}$ & $1\times10^{-4}$ \\

\hline
\end{tabular}
\caption{Representative physical parameters and initial conditions used as input for the fourth--order Runge--Kutta (RK4) simulations of the condensate width dynamics for $^{23}$Na.}
\label{Table4}
\end{table}

\begin{table}[h]
\centering
\begin{tabular}{c c c}
\hline
Parameter & Symbol & Value \\
\hline
Scattering length & $a_s$ & $-1.45\times10^{-9}$ m \\
Interaction strength & $g=4\pi\hbar^2 a_s/m$ & $-1.7\times10^{-50}$ J m$^3$ \\
\hline
\end{tabular}
\caption{Physical constants for $^{7}$Li used in the simulations.}
\label{Table5}
\end{table}

\begin{table}[h]
\centering
\begin{tabular}{c c c c c c}
\hline
Case & $N$ & $\lambda$ (J m$^3$) & $\beta$ (J) & $a(0)$ (m) & $\dot a(0)$ (m/s) \\
\hline
A & $8\times10^3$ & $-1.7\times10^{-50}$ & $5\times10^{-33}$ & $3\times10^{-6}$ & $0$ \\

B & $1.5\times10^4$ & $-1.7\times10^{-50}$ & $1\times10^{-32}$ & $2.5\times10^{-6}$ & $0$ \\

C & $3\times10^4$ & $-1.7\times10^{-50}$ & $2\times10^{-32}$ & $2\times10^{-6}$ & $0$ \\

D & $5\times10^4$ & $-1.7\times10^{-50}$ & $3\times10^{-32}$ & $1.5\times10^{-6}$ & $1\times10^{-4}$ \\

\hline
\end{tabular}
\caption{Representative physical parameters and initial conditions used as input for the fourth--order Runge--Kutta (RK4) simulations of the condensate width dynamics for $^{7}$Li.}
\label{Table6}
\end{table}
\subsection{Dimensionless Rescaling of the Width Equation}

For realistic atomic parameters of Bose--Einstein condensates, the dynamical
equation for the condensate width contains coefficients with very different
orders of magnitude. This may cause numerical instabilities when integrating
the equation using explicit methods such as the fourth--order Runge--Kutta
scheme. To avoid this issue, it is convenient to introduce dimensionless
variables.

We define a characteristic length scale $a_*$ and a characteristic time scale
$t_*$, and introduce the rescaled variables

\begin{equation}
a(t) = a_* x(\tau), \qquad t = t_* \tau .
\end{equation}

Derivatives transform as

\begin{equation}
\dot a = \frac{a_*}{t_*} x', \qquad
\ddot a = \frac{a_*}{t_*^2} x'' ,
\end{equation}

where primes denote derivatives with respect to the dimensionless time
$\tau$.

Substituting these expressions into the dynamical equation

\begin{equation}
\ddot a
=
\frac{\hbar^2}{m^2 a^3}
+
\frac{Q^2}{a^3}
-
\frac{m^2 c^4}{\hbar^2} a
-
\frac{3\lambda N}{2(2\pi)^{3/2} a^4}
+
\frac{3\beta c^2}{a},
\end{equation}

we obtain

\begin{equation}
\frac{a_*}{t_*^2} x''
=
\frac{\hbar^2}{m^2 a_*^3}\frac{1}{x^3}
+
\frac{Q^2}{a_*^3}\frac{1}{x^3}
-
\frac{m^2 c^4}{\hbar^2} a_* x
-
\frac{3\lambda N}{2(2\pi)^{3/2} a_*^4}\frac{1}{x^4}
+
\frac{3\beta c^2}{a_*}\frac{1}{x}.
\end{equation}

Multiplying by $t_*^2/a_*$ yields the dimensionless equation

\begin{equation}
x'' =
\frac{\kappa_1}{x^3}
+
\frac{\kappa_2}{x^3}
-
\kappa_3 x
-
\frac{\kappa_4}{x^4}
+
\frac{\kappa_5}{x},
\label{EQ:0001}
\end{equation}

where the dimensionless parameters are

\begin{align}
\kappa_1 &= \frac{\hbar^2 t_*^2}{m^2 a_*^4}, \\
\kappa_2 &= \frac{Q^2 t_*^2}{a_*^4}, \\
\kappa_3 &= \frac{m^2 c^4 t_*^2}{\hbar^2}, \\
\kappa_4 &= \frac{3\lambda N t_*^2}{2(2\pi)^{3/2} a_*^5}, \\
\kappa_5 &= \frac{3\beta c^2 t_*^2}{a_*^2}.
\end{align}

The choice of $a_*$ and $t_*$ can be made so that some coefficients become
of order unity, significantly improving the numerical stability of the
integration. A natural relativistic choice is $t_*=\hbar/(mc^2)$, which yields $\kappa_3=1$. Taking a characteristic condensate size $a_* = 1\,\mu\mathrm{m}$, typical parameters of a Bose--Einstein condensate lead to the approximate orders of magnitude $
\kappa_1 \sim 10^{-46},
\kappa_2 \sim 10^{-30},
\kappa_3 = 1,
\kappa_4 \sim 10^{-68},
\kappa_5 \sim 10^{-55}. $

With this rescaling the dominant contribution becomes the linear term $-x$, while the remaining terms encode quantum pressure, cubic interactions, and the logarithmic correction. This dimensionless form is therefore convenient for numerical integration using standard methods such as Runge--Kutta. For completeness, it is useful to analyze the linear part of the dimensionless equation. 
If only the dominant term proportional to $\kappa_3$ is retained, the equation reduces to

\[
x'' = -\kappa_3 x,
\]

which corresponds to the equation of a simple harmonic oscillator. The general analytical solution is

\[
x(\tau)=C_1\cos(\sqrt{\kappa_3}\,\tau)+C_2\sin(\sqrt{\kappa_3}\,\tau),
\]

where $C_1$ and $C_2$ are determined by the initial conditions. For $x(0)=x_0$ and 
$x'(0)=v_0$, the solution becomes

\[
x(\tau)=x_0\cos(\sqrt{\kappa_3}\,\tau)+\frac{v_0}{\sqrt{\kappa_3}}\sin(\sqrt{\kappa_3}\,\tau).
\]

In the particular case where the characteristic time scale is chosen as 
$t_*=\hbar/(mc^2)$, one obtains $\kappa_3=1$, and the equation simplifies to

\[
x''=-x,
\]

with solution

\[
x(\tau)=x_0\cos\tau + v_0\sin\tau.
\]

\begin{figure}
    \centering
   \includegraphics[width=1\linewidth]{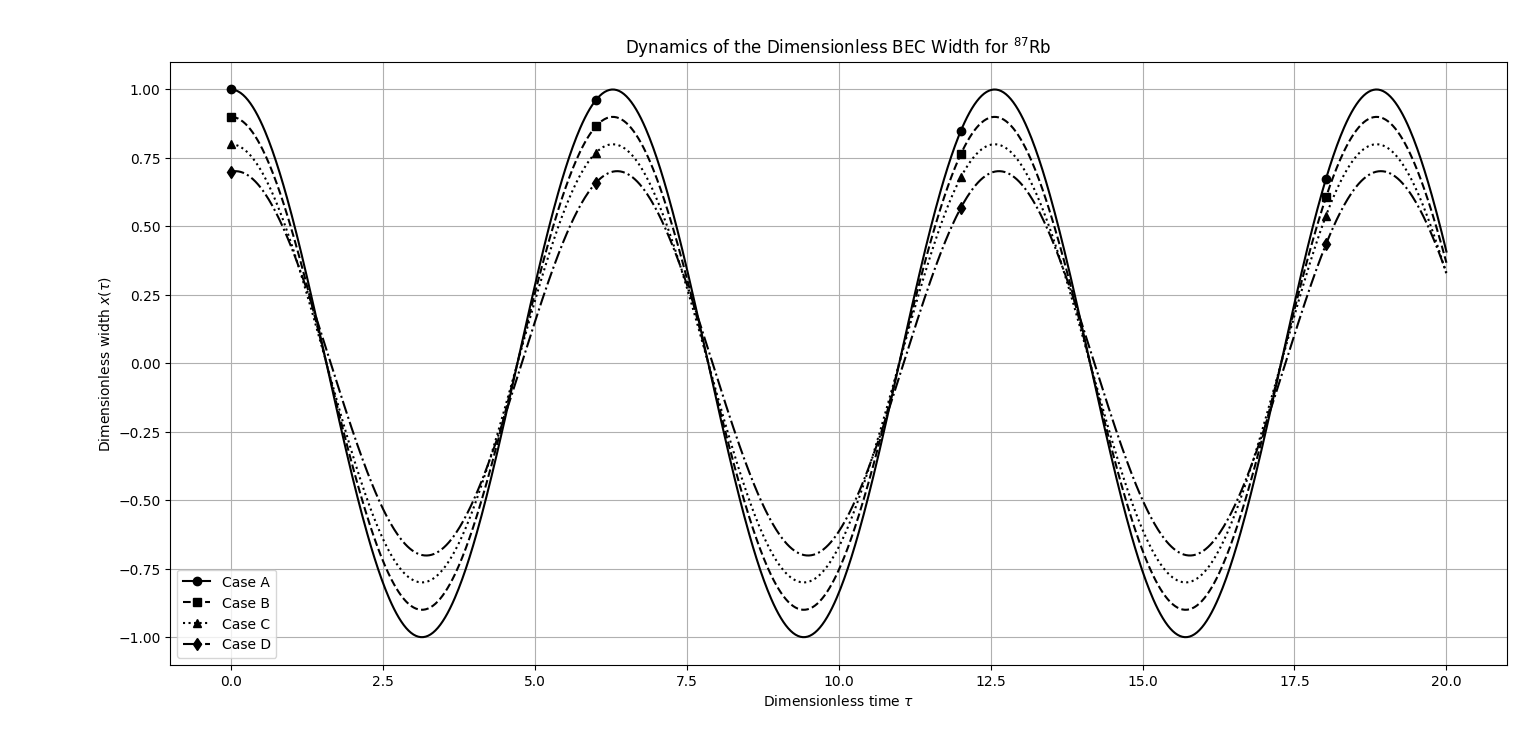}
    \caption{Numerical solutions of Eq. \ref{EQ:0001} were obtained using the parameter sets listed in Tables \ref{Table1} and \ref{Table2} for rubidium, Tables \ref{table3} and \ref{Table4} for sodium, and Tables \ref{Table5} and \ref{Table6} for lithium. Despite the differences in atomic mass and corresponding dimensionless coefficients, all three species exhibit qualitatively similar dynamical behavior. In particular, the time evolution of the condensate width shows regular oscillatory patterns, indicating that the dominant contribution to the dynamics is governed by the effective harmonic term in the rescaled equation of motion.}
    \label{FIG: Normal}
\end{figure}

This result shows that, in the absence of nonlinear contributions, the condensate width 
performs harmonic oscillations in the dimensionless time variable $\tau$. The remaining 
terms in the full equation, therefore act as nonlinear corrections to this fundamental 
oscillatory dynamics, as shown in Fig. \ref{FIG: Normal}. The consistency across different atomic species suggests that the system dynamics are largely controlled by the structure of the effective equation rather than the specific microscopic parameters. Variations in mass primarily affect the scaling of the dimensionless coefficients, slightly modifying the amplitude and frequency of oscillations, but do not alter the overall qualitative behavior. This robustness supports the validity of the rescaling procedure and indicates that the observed self-confined oscillatory regime is a generic feature of the model within the explored parameter range.

These results also reinforce the interpretation that the balance between the effective restoring force and the nonlinear interaction terms leads to bounded dynamics, preventing both collapse and unbounded expansion under the chosen conditions.
\subsection{Alternative Rescaling}

In order to improve the numerical behavior of the equation and bring the dominant
physical contributions to comparable magnitude, it is convenient to introduce
a different set of characteristic scales. Starting from the equation for the
condensate width

\[
\ddot a =
\frac{\hbar^2}{m^2 a^3}
+
\frac{Q^2}{a^3}
-
\frac{m^2 c^4}{\hbar^2} a
-
\frac{3\lambda N}{2(2\pi)^{3/2} a^4}
+
\frac{3\beta c^2}{a},
\]

we define dimensionless variables

\[
a = a_* x, \qquad t = t_* \tau .
\]

A convenient choice for the characteristic length scale is obtained by
balancing the quantum pressure term with the harmonic contribution,

\[
\frac{\hbar^2}{m^2 a^3} \sim \frac{m^2 c^4}{\hbar^2} a,
\]

which yields

\[
a_*^4 = \frac{\hbar^4}{m^4 c^4}.
\]

Therefore,

\[
a_* = \frac{\hbar}{mc}.
\]

For the time scale we take the natural relativistic value

\[
t_* = \frac{\hbar}{mc^2}.
\]

Introducing these variables into the equation of motion and dividing by the
appropriate factors leads to the dimensionless equation

\begin{equation}
x'' =
\frac{1}{x^3}
+
\tilde{\kappa}_2 \frac{1}{x^3}
-
x
-
\frac{\tilde{\kappa}_4}{x^4}
+
\frac{\tilde{\kappa}_5}{x},
\label{EQ:0002}
\end{equation}

where the dimensionless parameters are

\[
\tilde{\kappa}_2 =
\frac{Q^2 m^2 c^2}{\hbar^4},
\]

\[
\tilde{\kappa}_4 =
\frac{3\lambda N m^3 c^3}{2(2\pi)^{3/2}\hbar^3},
\]

\[
\tilde{\kappa}_5 =
\frac{3\beta}{mc^2}.
\]

With this choice of scales the dominant contributions become $\frac{1}{x^3} - x$,  which correspond to the competition between quantum pressure and the
effective harmonic term. The remaining terms encode the effects of the
conserved phase contribution, the cubic interaction, and the logarithmic
nonlinearity. This form is particularly convenient for numerical
integration since the leading terms are now of order unity.
\begin{figure}
    \centering
    \includegraphics[width=1\linewidth]{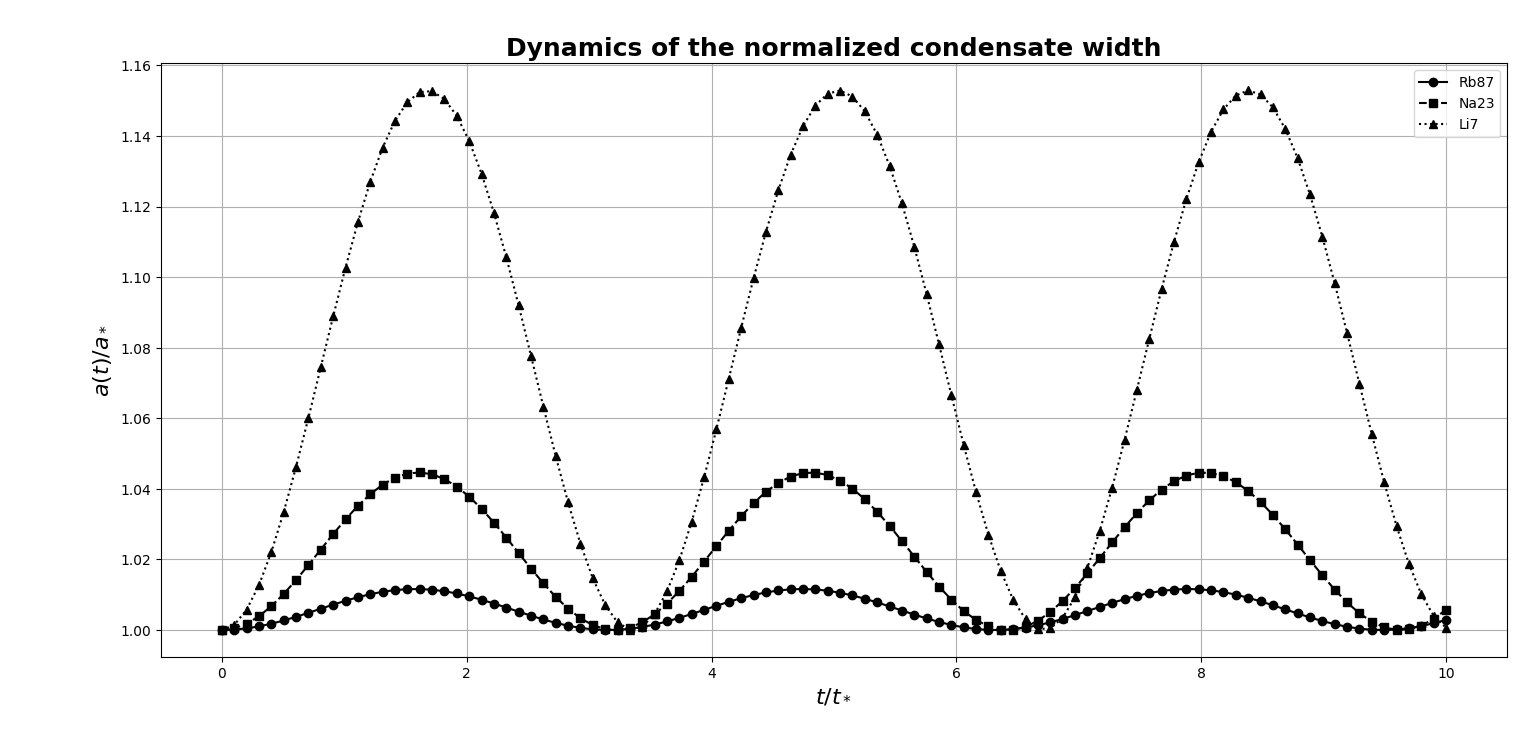}
    \caption{Numerical solutions for Eq. \ref{EQ:0002} using $Q=0, \lambda=0$ and $\beta=1\times10^{-10}$ with $N=1\times10^5$ atoms. }
    \label{FIG:Beta}
\end{figure}
\section{Discussion}

In this work we analyzed the time evolution of the width of a Bose–Einstein condensate described by a relativistic nonlinear equation containing both cubic and logarithmic interaction terms. By applying a Gaussian variational ansatz, the original field equation was reduced to an effective ordinary differential equation governing the condensate width $a(t)$. This approach allows the complex field dynamics to be interpreted in terms of a single collective coordinate representing the size of the condensate.

\begin{figure}
    \centering
    \includegraphics[width=1\linewidth]{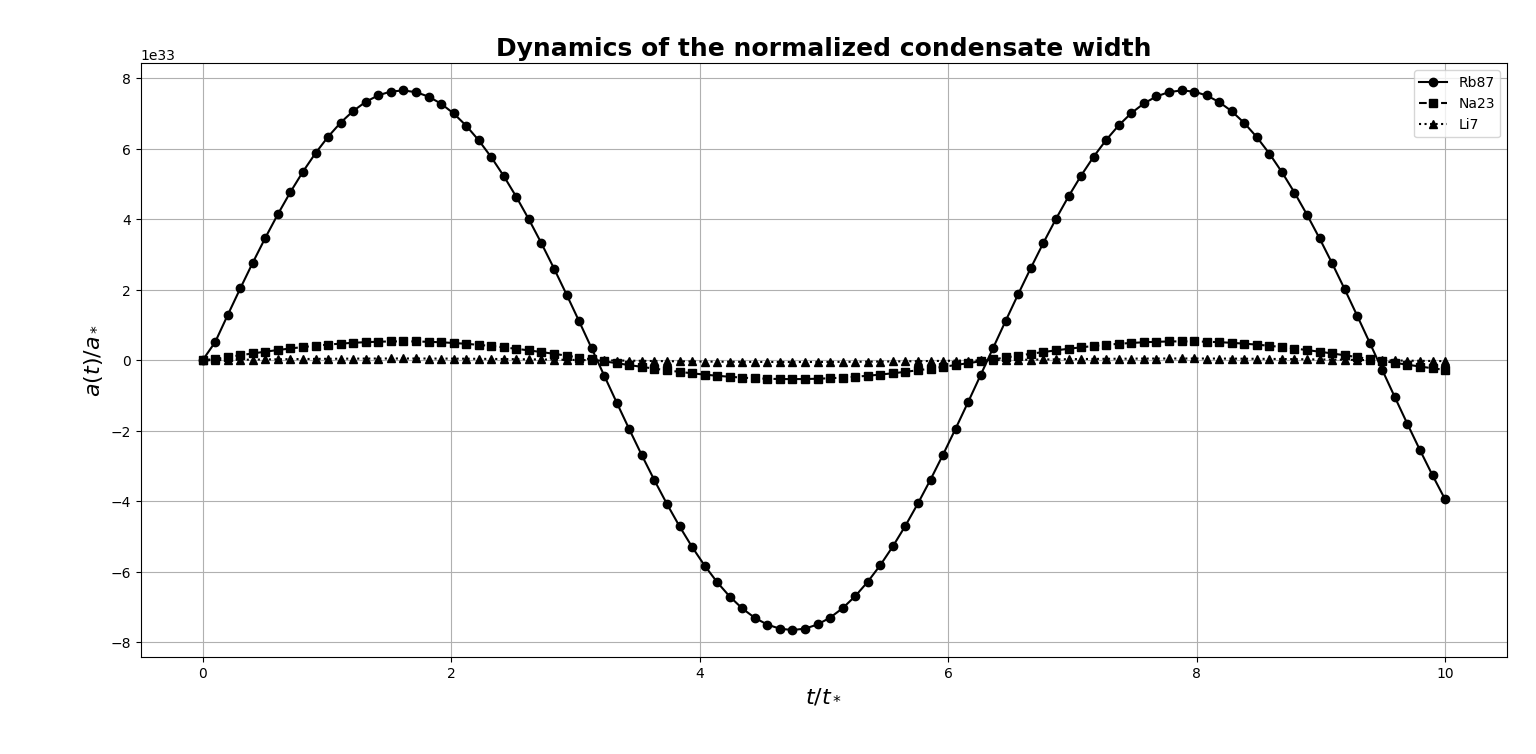}
    \caption{Numerical solutions for Eq. \ref{EQ:0002} using $Q=1\times 10^{-34}, \lambda=0$ and $\beta=0$ with $N=1\times10^5$ atoms. }
    \label{FIG:Q}
\end{figure}

The resulting equation contains several competing physical contributions. The term proportional to $1/a^{3}$ originates from the quantum kinetic pressure associated with the uncertainty principle, while the linear restoring term proportional to $a$ arises from the relativistic mass contribution. The cubic interaction produces an additional nonlinear term proportional to $1/a^{4}$, whose sign depends on the interaction parameter $\lambda$. For attractive interactions ($\lambda<0$) this term tends to compress the condensate and may lead to collapse if no additional stabilizing mechanism is present.

An important feature of the model is the presence of the logarithmic nonlinearity. This term introduces a contribution proportional to $1/a$ in the effective equation of motion and acts as an effective pressure that can counterbalance attractive interactions. Such logarithmic contributions have been discussed in several contexts, including effective descriptions of quantum droplets and generalized mean–field theories. In the present model, the logarithmic term can modify the stability of the condensate and shift the equilibrium radius.
\begin{figure}
    \centering
   \includegraphics[width=1\linewidth]{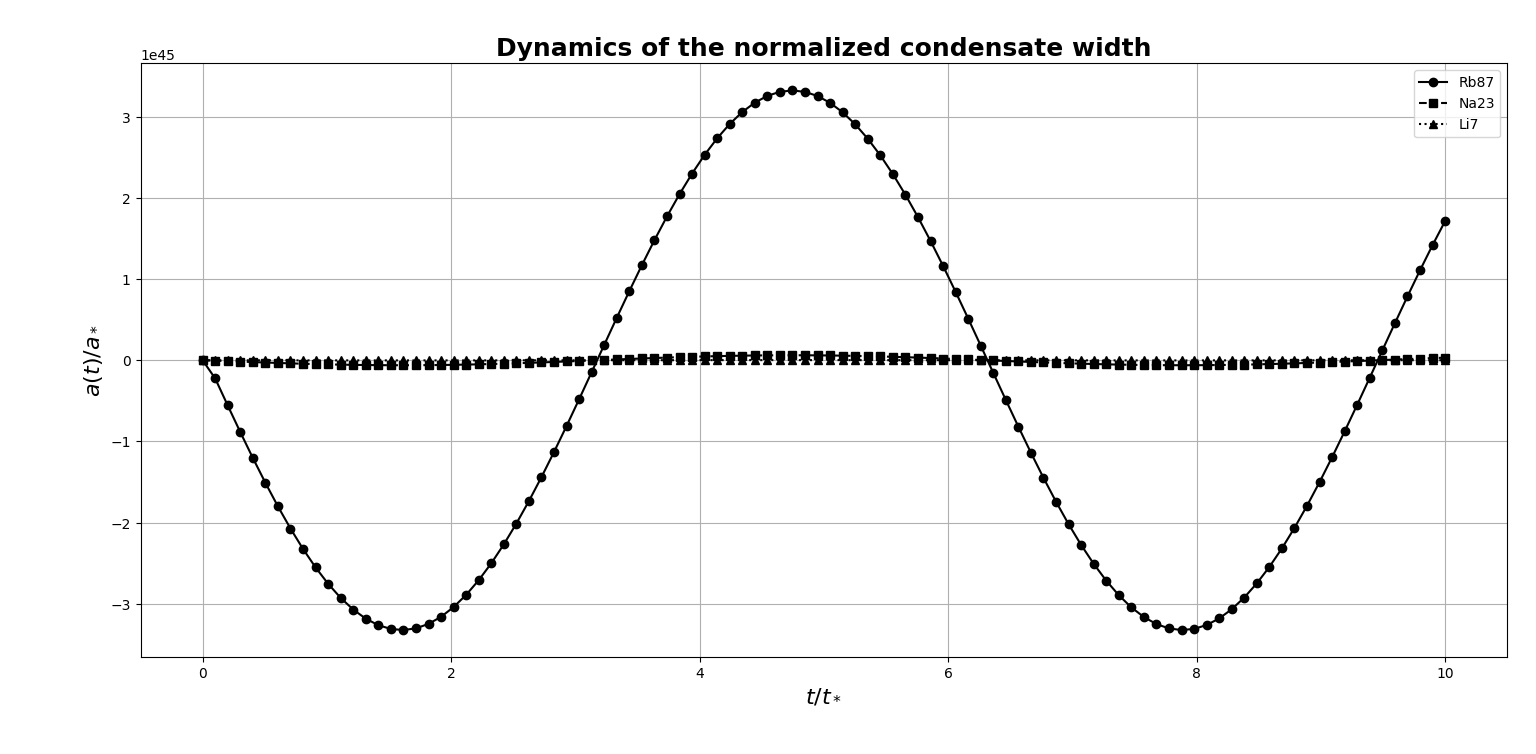}
    \caption{Numerical solutions for Eq. \ref{EQ:0002} using $Q=0, \lambda=1\times10^{-10}$ and $\beta=0$ with $N=1\times10^5$ atoms. }
    \label{FIG:Lambda}
\end{figure}

After introducing suitable dimensionless variables, the equation of motion was integrated numerically using a fourth–order Runge–Kutta scheme. Simulations were performed for parameters corresponding to different atomic species commonly used in experiments, including rubidium, sodium, and lithium. The results show that the dominant contribution to the dynamics is typically the harmonic restoring term, which leads to oscillatory behavior of the condensate width around an equilibrium configuration. This explains the approximately sinusoidal profiles observed in the numerical solutions.
\begin{figure}
    \centering
    \includegraphics[width=1\linewidth]{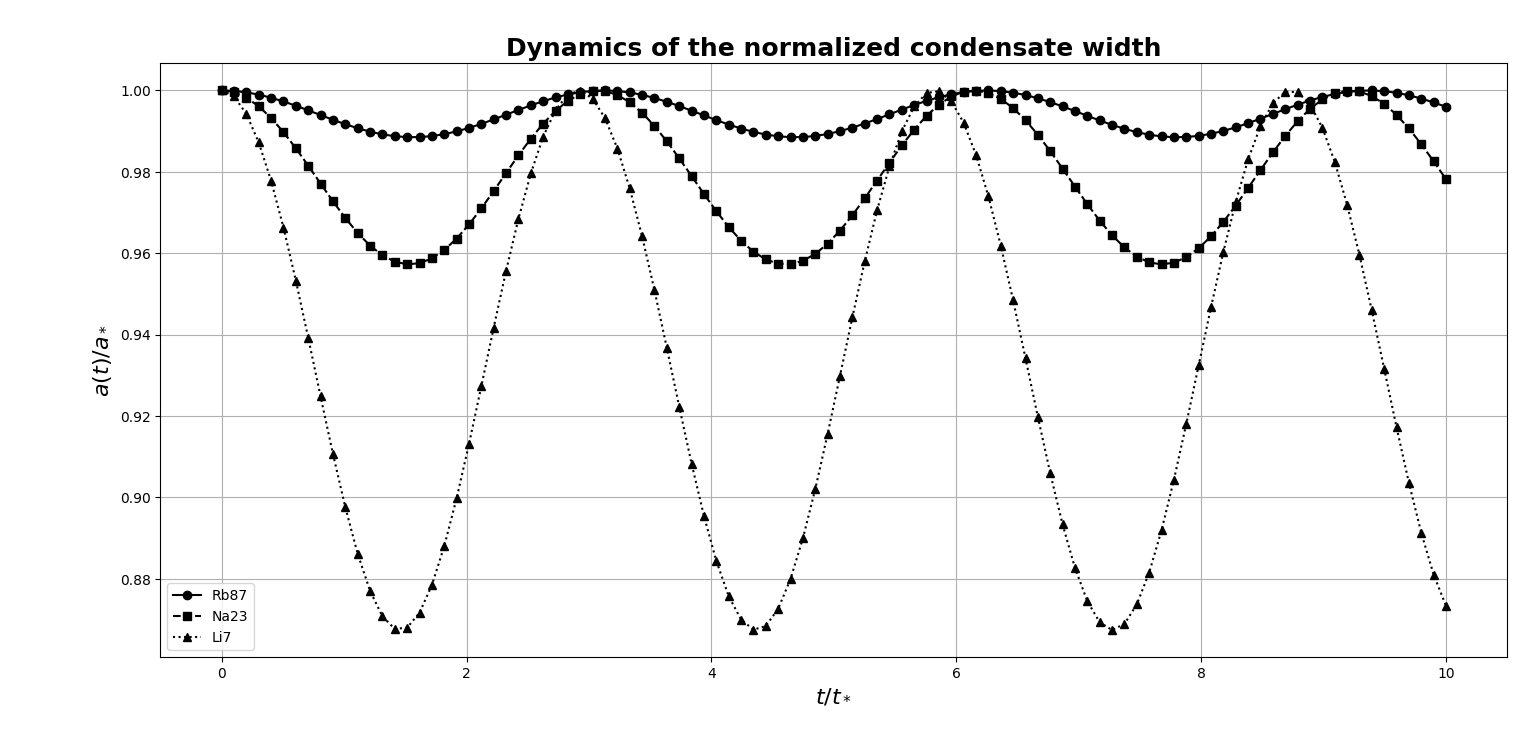}
    \caption{Numerical solutions for Eq. \ref{EQ:0002} using $Q=0, \lambda=0$ and $\beta=-1\times10^{-10}$ with $N=1\times10^5$ atoms. }
    \label{FIG:beta-}
\end{figure}
The differences between atomic species appear mainly through the mass dependence of the dimensionless parameters. Lighter atoms, such as lithium, produce larger effective nonlinear coefficients, which can enhance the influence of interaction terms and potentially lead to stronger deviations from harmonic behavior. In contrast, heavier atoms like rubidium tend to produce more regular oscillations dominated by the leading restoring term.
The numerical simulations presented in this work rely on a reduced dynamical description of the condensate based on a Gaussian variational ansatz. While this approach captures the essential collective behavior of the system, it inherently restricts the dynamics to a specific class of configurations and therefore does not represent the full solution space of the underlying field equation. In particular, phenomena such as fragmentation, vortex formation, or deviations from Gaussian profiles are not accounted for within this framework.

\begin{figure*}
    \centering
    \includegraphics[width=1\linewidth]{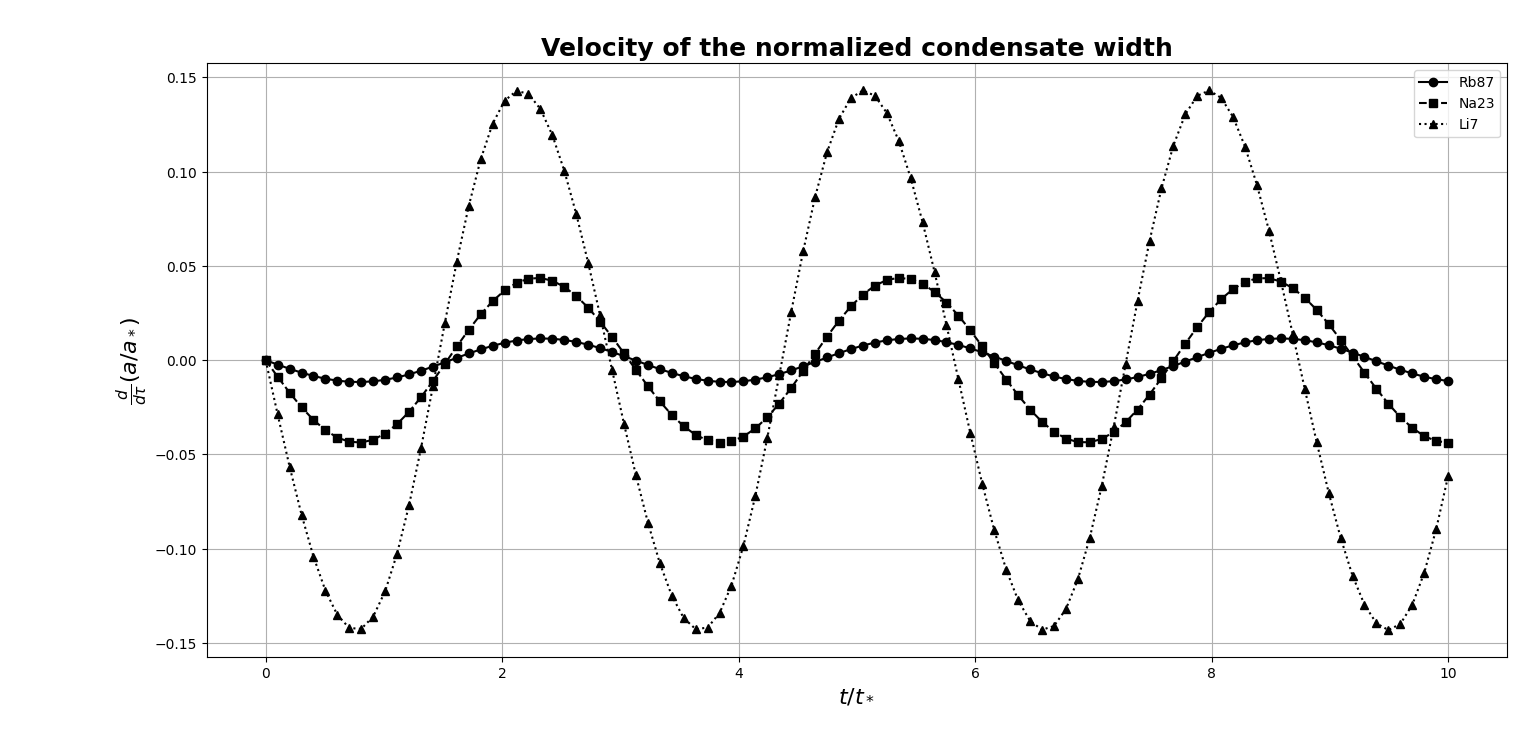}
    \caption{Free expansion velocity for Eq. \ref{EQ:0002} using $Q=0, \lambda=0$ and $\beta=-1\times10^{-10}$ with $N=1\times10^5$ atoms. }
    \label{FIG:beta1-}
\end{figure*}
The parameters $\lambda$, $\beta$, and $Q$ play a central role in determining the qualitative behavior of the system. In the present study, their values were chosen to ensure numerical stability while allowing exploration of distinct physical regimes. The cubic interaction strength $\lambda$ controls the mean-field contribution: negative values ($\lambda < 0$) correspond to attractive interactions that favor compression and may lead to collapse, whereas positive values promote expansion. The logarithmic parameter $\beta$ introduces an additional nonlinear contribution that effectively acts as a pressure term, which can counterbalance attractive interactions and stabilize the condensate. The conserved quantity $Q$, associated with the global phase symmetry, contributes an effective kinetic term that enhances repulsion at small scales.

The interplay between these parameters defines three main dynamical regimes. For sufficiently strong attractive interactions (large negative $\lambda$), the system tends toward collapse, characterized by a rapid decrease of the condensate width. When repulsive contributions dominate, either through $\beta$ or $Q$, the system exhibits expansion or dispersive behavior. Between these two extremes, a regime of self-confined oscillations emerges, where the condensate undergoes bounded, approximately periodic variations around an equilibrium radius. This regime is particularly relevant for the study of self-bound states and droplet-like configurations.

It is important to emphasize that the specific numerical values of the parameters were not chosen to exactly reproduce a particular experimental system, but rather to explore representative dynamical behaviors within a physically motivated range. A more precise quantitative comparison with experiments would require incorporating additional effects, such as external trapping potentials, finite-temperature corrections, and beyond-mean-field contributions. Nevertheless, the present model provides valuable qualitative insight into the mechanisms governing stability, collapse, and self-confinement in nonlinear relativistic condensates.
Overall, the variational approach provides a useful framework for studying the collective dynamics of condensates with generalized nonlinear interactions. Future work could explore regimes where the logarithmic contribution becomes dominant, investigate stability conditions for stationary solutions, or extend the analysis to include trapping potentials and dissipative effects.
\section{Conclusions}

In this work we investigated the dynamical evolution of the width of a Bose--Einstein condensate described by a relativistic nonlinear equation including both cubic and logarithmic interaction terms. By employing a Gaussian variational ansatz, the field dynamics were reduced to an effective ordinary differential equation governing the condensate width. This approach allowed the complex many-body dynamics to be interpreted in terms of a single collective variable representing the size of the condensate.

The resulting equation of motion contains several competing physical contributions, including quantum pressure, relativistic mass effects, cubic mean-field interactions, and a logarithmic nonlinear term. After introducing appropriate dimensionless variables, the equation was numerically integrated using a fourth-order Runge--Kutta method. This rescaling proved essential for improving numerical stability and for allowing a meaningful comparison between different atomic species.

Numerical simulations were performed for parameter sets corresponding to commonly studied condensates of rubidium, sodium, and lithium. The results show that the dominant contribution to the dynamics is typically the harmonic restoring term, leading to oscillatory behavior of the condensate width around an equilibrium configuration. This explains the approximately sinusoidal time evolution observed in the simulations. Differences between atomic species arise through the dependence of the dimensionless coefficients on the atomic mass, which modifies the relative importance of the nonlinear interaction terms.

The present analysis demonstrates that the variational approach provides a simple and effective framework for studying collective dynamics in condensates with generalized nonlinear interactions. In particular, the inclusion of the logarithmic term introduces an additional mechanism that can influence the stability and equilibrium properties of the system. The presence of oscillatory behavior in the condensate radius indicates that the system remains dynamically confined within a finite spatial region. Since the volume of the condensate scales as $V \sim a^3(t)$, bounded oscillations of the width imply that the condensate neither undergoes collapse nor expands indefinitely, but instead remains self-confined. This dynamical confinement is a key signature of self-bound states and supports the existence of quantum droplet-like configurations within the parameter regimes explored in this work. The persistence of these oscillations across different atomic species further suggests that such self-confined states are a robust feature of the model, arising from the balance between effective restoring forces and nonlinear interactions.

Future work may explore regimes where the nonlinear terms become dominant, investigate the existence and stability of stationary solutions, or extend the model to include external trapping potentials, dissipative effects, or higher-dimensional collective modes.

%
%






\bibliographystyle{plain}
\bibliography{apssamp1}

@article{RevModPhys.71.463,
  title = {Theory of Bose-Einstein condensation in trapped gases},
  author = {Dalfovo, Franco and Giorgini, Stefano and Pitaevskii, Lev P. and Stringari, Sandro},
  journal = {Rev. Mod. Phys.},
  volume = {71},
  issue = {3},
  pages = {463--512},
  numpages = {0},
  year = {1999},
  month = {Apr},
  publisher = {American Physical Society},
  doi = {10.1103/RevModPhys.71.463},
  url = {https://link.aps.org/doi/10.1103/RevModPhys.71.463}
}

@article{wynar2000molecules,
  title={Molecules in a Bose-Einstein condensate},
  author={Wynar, Roahn and Freeland, RS and Han, DJ and Ryu, C and Heinzen, DJ},
  journal={Science},
  volume={287},
  number={5455},
  pages={1016--1019},
  year={2000},
  publisher={American Association for the Advancement of Science}
}

@article{PhysRevLett.80.2027,
  title = {Optical Confinement of a Bose-Einstein Condensate},
  author = {Stamper-Kurn, D. M. and Andrews, M. R. and Chikkatur, A. P. and Inouye, S. and Miesner, H.-J. and Stenger, J. and Ketterle, W.},
  journal = {Phys. Rev. Lett.},
  volume = {80},
  issue = {10},
  pages = {2027--2030},
  numpages = {0},
  year = {1998},
  month = {Mar},
  publisher = {American Physical Society},
  doi = {10.1103/PhysRevLett.80.2027},
  url = {https://link.aps.org/doi/10.1103/PhysRevLett.80.2027}
}

@article{cabrera2018quantum,
  title={Quantum liquid droplets in a mixture of Bose-Einstein condensates},
  author={Cabrera, CR and Tanzi, L and Sanz, J and Naylor, B and Thomas, P and Cheiney, P and Tarruell, Leticia},
  journal={Science},
  volume={359},
  number={6373},
  pages={301--304},
  year={2018},
  publisher={American Association for the Advancement of Science}
}

@article{PhysRevLett.116.215301,
  title = {Observation of Quantum Droplets in a Strongly Dipolar Bose Gas},
  author = {Ferrier-Barbut, Igor and Kadau, Holger and Schmitt, Matthias and Wenzel, Matthias and Pfau, Tilman},
  journal = {Phys. Rev. Lett.},
  volume = {116},
  issue = {21},
  pages = {215301},
  numpages = {6},
  year = {2016},
  month = {May},
  publisher = {American Physical Society},
  doi = {10.1103/PhysRevLett.116.215301},
  url = {https://link.aps.org/doi/10.1103/PhysRevLett.116.215301}
}

@article{dong2020multi,
  title={Multi-stable quantum droplets in optical lattices},
  author={Dong, Liangwei and Qi, Wei and Peng, Ping and Wang, Linxue and Zhou, Hui and Huang, Changming},
  journal={Nonlinear Dynamics},
  volume={102},
  pages={303--310},
  year={2020},
  publisher={Springer}
}

@article{rodriguez2021oscillating,
  title={Oscillating Quantum Droplets From the Free Expansion of Logarithmic One-dimensional Bose Gases},
  author={Rodr{\'\i}guez-L{\'o}pez, Omar Abel and Castellanos, El{\'\i}as},
  journal={Journal of Low Temperature Physics},
  volume={204},
  number={3},
  pages={111--128},
  year={2021},
  publisher={Springer}
}

@article{avdeenkov2011quantum,
  title={Quantum Bose liquids with logarithmic nonlinearity: Self-sustainability and emergence of spatial extent},
  author={Avdeenkov, Alexander V and Zloshchastiev, Konstantin G},
  journal={Journal of Physics B: Atomic, Molecular and Optical Physics},
  volume={44},
  number={19},
  pages={195303},
  year={2011},
  publisher={IOP Publishing}
}

@book{pethick2008bose,
  title={Bose--Einstein condensation in dilute gases},
  author={Pethick, Christopher J and Smith, Henrik},
  year={2008},
  publisher={Cambridge university press}
}

@article{luo2021new,
  title={A new form of liquid matter: Quantum droplets},
  author={Luo, Zhi-Huan and Pang, Wei and Liu, Bin and Li, Yong-Yao and Malomed, Boris A},
  journal={Frontiers of Physics},
  volume={16},
  number={3},
  pages={32201},
  year={2021},
  publisher={Springer}
}

@article{astrakharchik2018dynamics,
  title={Dynamics of one-dimensional quantum droplets},
  author={Astrakharchik, GE and Malomed, Boris A},
  journal={Physical Review A},
  volume={98},
  number={1},
  pages={013631},
  year={2018},
  publisher={APS}
}

@article{visser2005massive,
  title={Massive Klein-Gordon equation from a Bose-Einstein-condensation-based analogue spacetime},
  author={Visser, Matt and Weinfurtner, Silke},
  journal={Physical Review D—Particles, Fields, Gravitation, and Cosmology},
  volume={72},
  number={4},
  pages={044020},
  year={2005},
  publisher={APS}
}

@article{castellanos2013klein,
  title={Klein--Gordon Fields and Bose--Einstein Condensates: Thermal Bath Contributions},
  author={Castellanos, El{\'\i}as and Matos, Tonatiuh},
  journal={International Journal of Modern Physics B},
  volume={27},
  number={11},
  pages={1350060},
  year={2013},
  publisher={World Scientific}
}

@article{megias2022nonlinear,
  title={Nonlinear Klein--Gordon equation and the Bose--Einstein condensation},
  author={Meg{\'\i}as, Eugenio and Teixeira, Marcio Jos{\'e} and Timoteo, Varese Salvador and Deppman, Airton},
  journal={The European Physical Journal Plus},
  volume={137},
  number={3},
  pages={325},
  year={2022},
  publisher={Springer}
}

@article{Zloshchastievql,
url = {https://doi.org/10.1515/zna-2017-0134},
title = {Stability and Metastability of Trapless Bose-Einstein Condensates and Quantum Liquids},
title = {},
author = {Konstantin G. Zloshchastiev},
pages = {677--687},
volume = {72},
number = {7},
journal = {Zeitschrift für Naturforschung A},
doi = {doi:10.1515/zna-2017-0134},
year = {2017},
lastchecked = {2025-06-17}
}

@article{zloshchastiev2012volume,
  title={Volume element structure and roton-maxon-phonon excitations in superfluid helium beyond the Gross-Pitaevskii approximation},
  author={Zloshchastiev, Konstantin G},
  journal={The European Physical Journal B},
  volume={85},
  number={8},
  pages={273},
  year={2012},
  publisher={Springer}
}

@article{Zloshchastievdilute,
author = {Zloshchastiev, Konstantin G.},
title = {Resolving the puzzle of sound propagation in a dilute Bose–Einstein condensate},
journal = {International Journal of Modern Physics B},
volume = {36},
number = {20},
pages = {2250121},
year = {2022},
doi = {10.1142/S0217979222501211},

URL = { 
    
        https://doi.org/10.1142/S0217979222501211
    
    

},
eprint = { 
    
        https://doi.org/10.1142/S0217979222501211
    
    

}
}

@article{mastache2024non,
  title={Non-Relativistic Boson Stars as Gravitational Quantum Droplets},
  author={Mastache, Jorge and Castellanos, El{\'\i}as and Chac{\'o}n-Acosta, Guillermo},
  journal={arXiv preprint arXiv:2411.01301},
  year={2024}
}

@article{hernandez2025rotating,
  title={Rotating Quantum Droplets in Low Dimensions},
  author={Hern{\'a}ndez, Kevin and Castellanos, El{\'\i}as},
  journal={Journal of Low Temperature Physics},
  pages={1--20},
  year={2025},
  publisher={Springer}
}

@article{gorka2009logarithmic,
  title={LOGARITHMIC KLEIN-GORDON EQUATION.},
  author={G{\'o}rka, Przemys{\l}aw},
  journal={Acta Physica Polonica B},
  volume={40},
  number={1},
  year={2009}
}

@article{bartkowski2008one,
  title={One-dimensional Klein--Gordon equation with logarithmic nonlinearities},
  author={Bartkowski, Konrad and G{\'o}rka, Przemys{\l}aw},
  journal={Journal of Physics A: Mathematical and Theoretical},
  volume={41},
  number={35},
  pages={355201},
  year={2008},
  publisher={IOP Publishing}
}

@article{morris1978classical,
  title={Classical theory of Klein--Gordon equations with logarithmic nonlinearities},
  author={Morris, TF},
  journal={Canadian Journal of Physics},
  volume={56},
  number={11},
  pages={1405--1411},
  year={1978},
  publisher={NRC Research Press Ottawa, Canada}
}

@article{fagnocchi2010relativistic,
  title={Relativistic Bose--Einstein condensates: a new system for analogue models of gravity},
  author={Fagnocchi, Serena and Finazzi, Stefano and Liberati, Stefano and Kormos, Marton and Trombettoni, Andrea},
  journal={New Journal of Physics},
  volume={12},
  number={9},
  pages={095012},
  year={2010},
  publisher={IOP Publishing}
}

@article{PhysRevLett.99.200406,
  title = {Bose-Einstein Condensation in the Relativistic Ideal Bose Gas},
  author = {Grether, M. and de Llano, M. and Baker, George A.},
  journal = {Phys. Rev. Lett.},
  volume = {99},
  issue = {20},
  pages = {200406},
  numpages = {4},
  year = {2007},
  month = {Nov},
  publisher = {American Physical Society},
  doi = {10.1103/PhysRevLett.99.200406},
  url = {https://link.aps.org/doi/10.1103/PhysRevLett.99.200406}
}

@article{10.1143/PTP.115.1047,
    author = {Fukuyama, Takeshi and Morikawa, Masahiro},
    title = {The Relativistic Gross-Pitaevskii Equation and Cosmological Bose-Einstein Condensation: Quantum Structure in the Universe},
    journal = {Progress of Theoretical Physics},
    volume = {115},
    number = {6},
    pages = {1047-1068},
    year = {2006},
    month = {06},
    issn = {0033-068X},
    doi = {10.1143/PTP.115.1047},
    url = {https://doi.org/10.1143/PTP.115.1047},
    eprint = {https://academic.oup.com/ptp/article-pdf/115/6/1047/5257462/115-6-1047.pdf},
}

@article{Fujita1991,
  author    = {Fujita, S. and Kimura, T. and Zheng, Y.},
  title     = {On the Bose-Einstein condensation of free relativistic bosons with or without mass},
  journal   = {Foundations of Physics},
  year      = {1991},
  volume    = {21},
  number    = {9},
  pages     = {1117--1130},
  doi       = {10.1007/BF00733389},
  url       = {https://doi.org/10.1007/BF00733389}
}

@article{Böhmer_2007,
doi = {10.1088/1475-7516/2007/06/025},
url = {https://doi.org/10.1088/1475-7516/2007/06/025},
year = {2007},
month = {jun},
publisher = {},
volume = {2007},
number = {06},
pages = {025},
author = {Böhmer, C G and Harko, T},
title = {Can dark matter be a Bose–Einstein condensate?},
journal = {Journal of Cosmology and Astroparticle Physics}
}

\end{document}